# A Hybrid Estimation of Distribution Algorithm with Random Walk local Search for Multi-mode Resource-Constrained Project Scheduling problems

**Omar S. Soliman [1], Elshimaa A. R. Elgendi [2]**

[1, 2] *(Faculty of Computers and Information, Cairo University, 5 Ahmed Zewail Street, Orman, 12613 Giza, Egypt.)*

***ABSTRACT :** Multi-mode resource-constrained project scheduling problems (MRCPSPs) are classified as NP-hard problems, in which a task has different execution modes characterized by different resource requirements. Estimation of distribution algorithm (EDA) has shown an effective performance for solving such real-world optimization problems but it fails to find the desired optima. This paper integrates a novel hybrid local search technique with EDA to enhance their local search ability. The new local search is based on delete-then-insert operator and a random walk (DIRW) to enhance exploitation abilities of EDA in the neighborhoods of the search space. The proposed algorithm is capable to explore and exploit the search mechanism in the search space through its outer and inner loops. The proposed algorithm is tested and evaluated using benchmark test problems of the project scheduling problem library PSPLIB. Simulation results of the proposed algorithm are compared with the classical EDA algorithm. The obtained results showed that the effectiveness of the proposed algorithm and outperformed the compared EDA algorithm.*

## 1. Introduction

The multi-mode resource-constrained project scheduling problem (MRCPSP) is a well-known optimization problem in the project scheduling literature. The MRCPSP is a generalized version of the single-mode resource constrained project scheduling problem (RCPSP). Generally, the MRCPSP is much closer to reality. In a multi-mode resource-constrained project, each activity has several execution modes. Each mode has related duration and nonrenewable and renewable resource requirements. The MRCPSP is more complex than the RCPSP. Kolisch and Drexl [1] have proven that even finding a feasible solution of the MRCPSP is NP-complete if there is more than one non-renewable resource. Therefore, heuristics for the MRCPSP have gained increasing attention during the past decades.

Hartmann [2] developed a GA to solve the MRCPSP, in which single-pass improvement and multi-pass improvement were adopted to enhance the local search. Alcaraz et al. [3] also proposed a GA to solve the MRCPSP. Different from the GA developed by Hartmann, a forward/backward gene was adopted in the solution representation, and a new fitness function was developed to improve the performance of the GA. Józefowska et al. [4] proposed a SA algorithm to solve the MRCPSP, in which two versions of SA were discussed: SA without penalty function and SA with penalty function. In both cases, three neighborhood generation mechanisms were applied: neighborhood shift, mode change and combination of neighborhood shift and mode change. Bouleimen and Lecocq [5] proposed a SA algorithm for both the RCPSP and the MRCPSP. For the MRCPSP, they used the activity list and mode list to represent a solution, and they also introduced two embedded search loops alternating activity and mode neighborhood exploration. Recently, Jarboui et al. [6] proposed a combinatorial PSO approach, in which a unique representation was adopted.

In addition, Van Peteghem and Vanhoucke [7] proposed an artificial immune system (AIS) to solve the MRCPSP. In the AIS, the mode assignment list generation was translated to a multi-choice multi-dimensional knapsack problem. Wauters et al. [8] proposed a multi-agent learning approach, where an agent was placed each activity node to decide the order to visit its successors and the mode to execute the activity. Damak et al. [9] proposed a differential evolution (DE) approach to solve the MRCPSP. Elloumi and Fortemps [10] proposed a hybrid rank-based evolutionary algorithm that transformed the MRCPSP to a bi-objective problem to cope with the potential violation of the nonrenewable resource constraints. Ranjbar et al. [11] proposed a hybrid scatter search to solve the discrete time/resource trade-off problem and the MRCPSP. Tseng and Chen [12] proposed a two-phase genetic local search algorithm that contained two phases: the first phase aimed at searching globally for promising areas while the second phase aimed at searching in the promising areas more thoroughly. Lova et al. [13] proposed an efficient hybrid GA to solve the MRCPSP, in which a unique mode assignment procedure was developed and a new fitness function was proposed to improve the results. Van Peteghem and Vanhoucke [14] proposed a bi-population GA to solve preemptive and non-preemptive MRCPSP that made use of two separate populations and extended the serial schedule generation scheme by introducing a mode improvement procedure. Wang and Fang [15] showed the effectiveness of using shuffled frog-leaping algorithm (SFLA) for solving the MRCPSP with the criterion to minimize the makespan. But SFLA needs a very careful choice of its parameter in order to be effective. Wang and Fang [16] designed an estimation of distribution algorithm (EDA) for solving the MRCPSP by using an encoding scheme based on the activity-mode list and a decoding scheme based on the multi-mode serial schedule generating scheme.

Estimation of distribution algorithm (EDA) is a kind of stochastic population-based optimization algorithm based on statistical learning [17]. So far, EDA has been developed to solve a variety of optimization problems in academic and engineering fields. In this paper, we proposed a hybrid EDA and with random walk as a local search method for solving nonpreemptive MRCPSP minimizing the project makespan. The new local search heuristic (DIRW) based on delete-then-insert operator and random walk is developed to enhance the exploitation ability of the algorithm. The rest of this paper is organized as follows; Section 2 describes the MRCPSPs. Section 3 introduces the proposed hybrid EDA algorithm for MRCPSP. In Sections 4 experimental results are presented, where the last section is devoted to conclusions.

## 2. Multi-Mode Resource-Constrained Project Scheduling Problem

In MRCPSP, the target is to study how to allocate renewable/nonrenewable resource and schedule activities to minimize the whole project makespan. Formally, we can describe the MRCPSP as follows. A project consists in $J$ activities, with a dummy start node 0 and a dummy end node $J + 1$. The precedence relations between activities are defined by a directed acyclic graph $G$. No activity may be started before all its predecessors are finished. Graph $G$ is numerically numbered, i.e. an activity has always a higher number than all its predecessors. $A$ is the set of all activities pairs $(A_i, A_j)$ such that $A_i$ directly precedes $A_j$. Each activity $j$, $j = 1, \cdots, J$ has to be executed in one of $M_j$ modes. The activities are non-preemptable and a mode chosen for an activity may not be changed (i.e. an activity $j$, $j = 1, \cdots, J$, started in mode, $m \in 1 \cdots M_j$ must be completed in mode $m$ without preemption). The duration of activity $j$ executed in mode $m$ is $d_{jm}$. We assume that there are R renewable and N non-renewable resources. The number of available units of renewable resource $k$, $k = 1, \cdots R$ is $R_k^\rho$ and the number of available units of non-renewable resource $l$, $l = 1, \cdots, N$, is $R_l^v$. Each activity j executed in mode m requires for its processing $r_{jmk}^\rho$ units of renewable resource $k$, $= 1, \cdots R$, and consumes $r_{jml}^v$ units of non-renewable resource $l$, $l = 1, \cdots, N$. We assume that all activities and resources are available at the beginning of the process. The objective of the MRCPSP is to find an assignment of modes to activities as well as precedence and resource-feasible starting times for all activities, such that the makespan of the project is minimized. The mathematical formulation of the MRCPSP found in Talbot [18].

## 3. The Proposed Hybrid EDA for MRCPSP

3.1 Proposed Algorithm Mechanism

The proposed algorithm hybrid EDA to solve the MRCPSP uses activity-mode list (AML)

decoding scheme described in section 3.2 below. The proposed algorithm applies a preprocessing reduction procedure in project data to efficiently reduce the search space, flowed by initializing the probability matrix uniformly distributed. Then generate an initial population of P individuals by sampling the probability matrix using probability generating mechanism. Then, all individuals are evaluated by the multi-mode serial schedule generation scheme (MSSGS). Next, best_P individuals are selected from the population by ranking selection. Then the Multi-mode double-justification (MDJ) is applied only to the selected best_P individuals to improve the makespan value. After that, a newly developed local search method (DIRW) is applied to the best individuals to exploit the neighborhood of the selected individuals. After the local search operation, the probability matrixes Pact(t) and Pmod(t) are updated based on the selected best_P individuals. This procedure is repeated until the stopping criterion is reached. Straight forwardly, the pseudo code of the proposed algorithm is described in Algorithm 1.

**Algorithm 1:** The hybrid EDA with DIRW local search for solving MRCPSP

1. Set t = 0;
2. Initialize the probability matrixes $P_{act}(0)$ and $P_{mod}(0)$;
3. While stopping criterion is not met do
    3.1. Generate the new population with P individuals by sampling the probability matrix using a permutation-based probability generating mechanism (PGM);
    3.2. Evaluate each individual using MSSGS;
    3.3. Select *best_P* individuals based on ranking selection;
    3.4. Apply MDJ to the selected best_P individuals;
    3.5. Apply Local search (DIRW) to the selected *best_P* individuals;
    3.6. t=t+1;
    3.7. Update the probability matrixes $P_{act}(t)$ and $P_{mod}(t)$;
4. Return the best found solution.

3.2    Solution Representation and decoding scheme

In our proposed algorithm, The encoding scheme chosen for individuals is an activity-mode list (AML), which consists of two vectors: (1) a precedence feasible activity list (AL) $\{A_{\pi_1}, \cdots, A_{\pi_i}, \cdots, A_{\pi_J}\}$; (2) a mode assignment list (ML) $\{m_{\pi_1}, \cdots, m_{\pi_i}, \cdots, m_{\pi_J}\}$. The $m_{\pi_i}$ in the ML indicates the mode of the $A_{\pi_i}$ in the AL. This representation makes it convenient and effective for solving the MRCPSP. That is the mode assignment list is needed to assign every activity a mode. The EDA does not operate on a schedule but on the AML representation of a schedule.

After the generation of population individuals, the serial schedule generation scheme (SGS) is used to evaluate the solution vectors, i.e., the solution vector of the MRCPSP is translated into a schedule [11]. We adopt the fitness function proposed by Lova et al. [13] to calculate fitness function for each individual.

3.3    Preprocessing

First of all, a preprocessing procedure is applied in project data to efficiently reduce the search space. The reduction procedure was first developed by Sprecher et al. [19] to exclude those modes which are inefficient or non-executable and those resources which are redundant. For more details we refer to Sprecher et al. [19], a mode is called inefficient if there is another mode of the same activity with the same or smaller duration and no more requirements for all resources. A mode is called non-executable if its execution would violate the renewable or nonrenewable resource constraints in any schedule. A nonrenewable resource is called redundant if the sum of the maximal requests for that nonrenewable resource does not exceed its availability. Excluding these modes or nonrenewable resources does not affect the set of feasible or optimal schedules.

3.4    Probability generating mechanism (PGM)

GA produces offspring through crossover and mutation operators. Not like the GA, EDA produces offspring by sampling according to a probability model, which captures and preserves the information found during the search, as well as any relevant heuristic information. One of the goals of probabilistic modeling in EDAs is to obtain a condensed, accurate model of the selected points.

In this regard, the permutation-based probability generating mechanism (PGM) developed by Wang and Fang [16] will be utilized. It is noted that EDA employs the same sampling method for generating both an initial population and a number of offspring populations over the generations during evolution.

### 3.5 Sample solution selection method

To select the sample solution for updating the probability model of the EDA, we employ the ranking selection as a countermeasure. The ranking selection selects the best individuals from the current population; i.e., the individuals with best fitness values.

### 3.6 Local search

After the best_P individuals are selected from the population, multi-mode double-justification (MDJ) is applied to the best_P individuals to improve their makespan values. MDJ iteratively employs the MSSGS forward and backward scheduling until no further improvement in the makespan of project can be found. More details of MDJ could be found in Wang and Fang [16].

To enhance the exploitation ability, we developed a new local search heuristic (DIRW) based on the delete-then insert operator and random walk. Delete-then-insert operator is applied on each activity on the AL which deletes the activity from its current position and then inserts it in a random eligible position. Accepting this move is based on acceptance rate (RW). DIRW enhance the exploitation ability by exploring the neighborhood of the individual as described in Algorithm 2.

The nonrenewable infeasible degree is calculated by (1). It is noticeable that the penalty $v_E(ML) = 0$ if the AML is nonrenewable resources feasible.

$$v_E(ML) = \sum_{l=1}^{N} \max\left\{0, \frac{\sum_{j=1}^{J} r_{jm_j l}^v - R_l^v}{R_l^v}\right\} \quad (1)$$

In the procedure of DIRW, we calculate nonrenewable infeasible degree $v_E(ML)$ first. If $v_E(ML) > 0$, assign the mode with minimum cumulative nonrenewable resources consumption to the currently modified activity; if $v_E(ML) = 0$, assign the mode with minimum duration to the currently modified activity.

**Algorithm 2**: The random walk Local search (DIRW)

```
Procedure DIRW Local Search(best_P)
Begin
i=0;
while(i<best_P) do
{
   i=i+1;
   AL'_i = AL_i;
   ML'_i = ML_i;
   //activity at first position is the start activity
which has no predecessor
   j=2;
   while(j<J) do
   {
     Generate uniformly distributed random
     number q, q ∈ [0,1]
     If(q < RW)
       Pos = position of Aj in AL'_i
       if (v_E(ML'_i) = 0)
           ML'_{i,pos}= min_m(d_{jm});
       else
         ML'_{i,pos} = min_m(∑_{l=1}^{N} r_{jml}^v);
       endif
       Delete Aj from position pos in ALi;
       Maxpos=max(π_k|(AL_{iπ_k}, A_j) ∈ A)+1;
       Minpos= min(π_k|(A_j, AL_{iπ_k}) ∈ A)-1;
       Insert Aj in a random location between
       Maxpos and Minpos on AL'_i;
       AL_i = AL'_i;
       ML_i = ML'_i;
     endif
     j=j+1
   }
   endwhile
}
endwhile
End
```

With the help of the proposed DIRW, when a new AML is generated (i.e. the AL and ML parts of the AML are changed by searching operation). In a word, the DIRW may reduce the nonrenewable resource consumption for infeasible representations and increase the nonrenewable resource consumption for feasible representations by selecting the modes with minimum duration.

## 4. Experimental Results
### 4.1 Test problems

In this section, we present the results of a computational experiment concerning the implementations of the proposed algorithm to solve MRCPSP. The experiments have been performed on an Intel(R) Celeron(R)-based DELL with 2.19 GHz clock-pulse and 1 Gb RAM. The proposed algorithm has been coded and compiled in MATLAB R2007b. We ran our program on standard MRCPSP test instances. These problems as well as their best found solutions are available in the project scheduling problem library PSPLIB generated using the project generator ProGen developed by Kolisch and Sprecher [20]. Optimal solutions for J10-20 instances sets are available, whereas no optimal solutions were found for J30 instances; hence, heuristic lower bounds are provided for the latter. In Table 1, we present the number of instances for which at least one feasible solution exists. Each instance of J10-20 and J30 contains three modes (i.e., $M_j = 3$, $j = 1, \cdots, J$; $M_0 = 1$; $M_{J+1} = 1$), two renewable and two nonrenewable resources. The duration of activity j in mode $m_j$ varies from 1 to 10.

Table1. Number of instances

| J | 10 | 12 | 14 | 16 | 18 | 20 | 30 |
|---|---|---|---|---|---|---|---|
| Number of Instances | 536 | 547 | 551 | 550 | 552 | 554 | 552 |

### 4.2 Configuration of the algorithm

In this section, we report the numerical configuration of the proposed algorithm through a numerical investigation.

The proposed algorithm contains four key parameters: the population size of each generation (P), the number of selected individuals to update the probability matrix (best_P), the learning speed ($\alpha$) and local search acceptance rate (*RW*). Following Wang and Fang [16], we settled on values of these parameters that are shown in Table 2.

The average relative deviation (ARD) value is the following average deviation value for N instances.

$$ARD = \frac{1}{N} \sum_{i=1}^{N} \frac{(Makespan_i - Optimum_i)}{Optimum_i} \quad (2)$$

where $Makespan_i$ is the makespan of the i[th] feasible instance obtained by the proposed algorithm; $Optimum_i$ is the optimal makespan of i[th] feasible instance.

Table2. Parameters values

|  | P | best_P | $\alpha$ | RW |
|---|---|---|---|---|
| Parameters value | 100 | 20%P | 0.5 | 0.5 |

We consider 5000/10000/20000/30000/40000/50000 generated schedules as stopping conditions. We run the proposed algorithm 20 times independently for each instance. The statistical results are listed in Table 3. Besides, we use Av.dev, optimum rate and feasibility rate for comparison. As for the lower bounds, the theoretically optimal values are used for set J10-20 and the critical-path based lower bounds are employed for set J30 since the theoretically optimal values are not available.

Table 3, shows the average deviation decreases constantly as the number of generated schedules increases for all the problem sets. It shows that our algorithm can keep finding better solutions as the population evolves which proves that the algorithm is very robust. From Table 3, similar conclusions can be drawn also about optimum and feasible rates.

To compare the proposed algorithm with the EDA algorithm developed by Wang and Fang [16], both algorithms have been coded and compiled in MATLAB R2007b. It is known that MATLAB compiler is too much slower than C++ compiler. So, results displayed in table 4 using 60 s CPU time as a stopping criterion. In Table 4, we compare the proposed algorithm and the EDA [16]. The stopping condition for the comparison is 60 s CPU time. From Table 4, it can be seen that the proposed algorithm not only has a higher optimal and feasible rates than EDA [16] but also has lower average deviation with less computation resource for all problem sets except J20 the classical EDA outperformed our proposed algorithm with a reasonable difference.

Table 3. The Av. dev, optimal rate and feasible rate for the proposed algorithm at different stopping conditions.

| Problem set | | Stopping condition | | | | | |
| --- | --- | --- | --- | --- | --- | --- | --- |
| | | 5000 | 10,000 | 20,000 | 30,000 | 40,000 | 50,000 |
| J10 | Av.dev (%) | 0.094 | 0.067 | 0.053 | 0.048 | 0.042 | 0.037 |
| | Optimal rate (%) | 97.629 | 98.175 | 98.442 | 98.542 | 98.672 | 98.756 |
| | Feasible rate (%) | 100 | 100 | 100 | 100 | 100 | 100 |
| J12 | Av.dev (%) | 0.120 | 0.103 | 0.069 | 0.057 | 0.049 | 0.041 |
| | Optimal rate (%) | 97.111 | 97.441 | 98.118 | 98.356 | 98.529 | 98.691 |
| | Feasible rate (%) | 100 | 100 | 100 | 100 | 100 | 100 |
| J14 | Av.dev (%) | 0.361 | 0.277 | 0.197 | 0.177 | 0.141 | 0.112 |
| | Optimal rate (%) | 92.313 | 93.985 | 95.572 | 95.983 | 96.693 | 97.266 |
| | Feasible rate (%) | 100 | 100 | 100 | 100 | 100 | 100 |
| J16 | Av.dev (%) | 0.424 | 0.324 | 0.250 | 0.196 | 0.189 | 0.173 |
| | Optimal rate (%) | 91.059 | 93.052 | 94.527 | 95.597 | 95.735 | 96.066 |
| | Feasible rate (%) | 100 | 100 | 100 | 100 | 100 | 100 |
| J18 | Av.dev (%) | 0.851 | 0.654 | 0.498 | 0.466 | 0.405 | 0.364 |
| | Optimal rate (%) | 82.461 | 86.437 | 89.580 | 90.230 | 91.428 | 92.251 |
| | Feasible rate (%) | 100 | 100 | 100 | 100 | 100 | 100 |
| J20 | Av.dev (%) | 1.093 | 0.946 | 0.811 | 0.758 | 0.705 | 0.672 |
| | Optimal rate (%) | 77.575 | 80.553 | 83.263 | 84.347 | 85.415 | 86.073 |
| | Feasible rate (%) | 100 | 100 | 100 | 100 | 100 | 100 |
| J30 | Av.dev (%) | 12.704 | 12.038 | 11.963 | 11.624 | 11.409 | 11.299 |
| | Optimal rate (%) | n/a | n/a | n/a | n/a | n/a | n/a |
| | Feasible rate (%) | 86.161 | 86.963 | 87.141 | 88.133 | 88.742 | 89.192 |

Note: n/a means not available, i.e., the optimal values of J30 are not available.

Table 4. Comparison with EDA [16] (60 s CPU time).

| J | Av.dev (%) | | Feasible rate (%) | | Optimal rate (%) | |
|---|---|---|---|---|---|---|
| | The proposed algorithm | EDA [16] | The proposed algorithm | EDA [16] | The proposed algorithm | EDA [16] |
| 10 | **0.053** | 0.067 | **100** | **100** | **98.4** | 98.2 |
| 12 | **0.071** | 0.086 | **100** | **100** | **98.1** | 97.8 |
| 14 | **0.200** | 0.247 | **100** | **100** | **95.5** | 94.6 |
| 16 | **0.330** | 0.403 | **100** | **100** | **92.9** | 91.5 |
| 18 | **0.500** | 0.548 | **100** | **100** | **89.5** | 88.6 |
| 20 | 0.810 | **0.771** | **100** | **100** | 83.3 | **84.1** |
| 30 | **12.01** | 13.90 | **87.01** | 80.1 | n/a | n/a |

Note: The bold values denote the best results among the results obtained by all the algorithms.

Next, the proposed algorithm has compared with the EDA [16] on the different set instances with 5000 computed schedules stopping criteria based on the PSPLIB to further show the effectiveness of the proposed algorithm. As reported in Table 5, the proposed algorithm has minimum average deviation over all test problems which proved that the exploitation ability of the proposed algorithm has been enhanced. The proposed algorithm outperforms the EDA [16] as EDA needs effort to find and track the most promising area at the beginning of the search procedure, as shown in Table 5.

Table 5. Comparison with EDA [16] (5000 computed schedules)

| Av.dev (%) | J10 | J12 | J14 | J16 | J18 | J20 |
|---|---|---|---|---|---|---|
| The proposed algorithm | **0.09** | **0.12** | **0.36** | **0.42** | **0.85** | **1.09** |
| EDA [16] | 0.12 | 0.14 | 0.43 | 0.59 | 0.90 | 1.28 |

Note: The bold values denote the best results among the results obtained by all the algorithms.

The comparisons between the proposed algorithm and EDA of Wang and Fang [16] showed that the proposed algorithm outperformed, and it is an effective algorithm for solving the MRCPSP.

## 5. Conclusion

This paper proposed a hybrid estimation of distribution algorithm with a new local search based on random walk and the delete-then-insert operator for solving the multi-mode resource-constrained project scheduling problems (MRCPSPs) to enhance the exploitation ability. The proposed algorithm utilizes the random walk and delete-then-insert operator as a local search method and multi-mode forward–backward improvement; to improve the exploitation ability of the proposed algorithm search space. The proposed algorithm is tested and evaluated using PSLIB. Experimental results and comparison with EDA [16] showed the effectiveness of the proposed algorithm. The further work is to develop a self-adaptive EDA with parameter learning mechanism and to extend the EDA method to solve other scheduling problems.